\documentclass[pra,twocolumn,tightenlines,showpacs,nofootinbib]{revtex4}
\usepackage{bm,dcolumn,amsmath,graphicx,amsfonts,amssymb}
\usepackage{epsfig}

\begin{document}
\title{High-precision atomic clocks with highly charged ions: nuclear
  spin-zero $f^{12}$-shell ions} 

\author{V.A. Dzuba$^{1,2}$, A. Derevianko$^1$,  and V. V. Flambaum$^2$}
\affiliation{
$^1$Department of Physics, University of Nevada, Reno, Nevada 89557, USA}
\affiliation{
$^2$School of Physics, University of New South Wales,
Sydney, NSW 2052, Australia}

\begin{abstract}
Optical atomic clocks using highly-charged ions hold an intriguing
promise of metrology 
at the 19th significant figure.
Here we study transitions within the $4f^{12}$ ground-state electronic 
configuration of  highly charged ions. We consider isotopes lacking
hyperfine structure and show that the detrimental effects of  coupling
of electronic quadrupole moments to gradients of trapping electric
field can be effectively reduced by using specially chosen virtual
clock transitions. 
The estimated systematic fractional clock accuracy is shown to be
below $10^{-19}$. 
\end{abstract}

\date{ \today }

\maketitle

%\section{Introduction}
Developing  accurate atomic clocks is important for both technological 
 and fundamental reasons. Cesium primary frequency standard which
is currently used to define the SI units of time and length  has
fractional accuracy  of 
the order $10^{-16}$~\cite{NIST}. 
State of the art clocks using trapped singly-charged ions  have
demonstrated fractional accuracies at the level of
$10^{-17}$~\cite{ChoHumKoe10}. 
Frequency standards based on neutral atoms trapped in optical lattice
aim at fractional 
accuracy of $10^{-18}$~\cite{Katori}. Further progress is possible with
clocks using nuclear optical transition~\cite{Th}, or clocks using optical
transitions in highly charged ions~\cite{HCI1,hole,Cf,IonClock,NdSm}. 

In our previous paper~\cite{IonClock} we proposed ion clocks based
on optical transitions in trapped highly charged ions (HCI). 
Clock HCI is co-trapped with lighter singly-charged ion (e.g., Be$^+$)
which is used for sympathetic cooling of the HCI and quantum-logic
clock readout and initialization. 
We identified HCIs with the $4f^{12}$ ground-state configuration to be
especially promising for precision time-keeping. It was demonstrated
that such ions can serve as a 
basis of a clockwork of exceptional accuracy, with fractional
uncertainty of about $10^{-19}$. One of the most
important systematic effect was determined to be the frequency shift
due to interaction 
of ionic quadrupole moments with gradients of trapping electric
field. It was suggested to use combinations of different hyperfine
transitions to suppress this shift.

Here we analyze similar $4f^{12}$ HCIs but propose another approach to
suppressing the quadrupole shift. Instead of using different hyperfine
transitions we combine transitions between states of different
projections of the total angular momentum. We focus on isotopes with
zero nuclear 
spin. Since these lack complicated hyperfine structure, the processes
of initializing the clock becomes easier. Also  the detrimental
second-order  AC Zeeman 
shift becomes substantially suppressed. In  the end, compared to the original 
proposal~\cite{IonClock},
our current scheme  can
 be easier to implement and  can have higher accuracy. 

The electronic states arising from the $4f^{12}$ configuration  have
some unique  
features which make them convenient when building very
accurate atomic clocks. First, transitions within these configuration are
always in the optical/near-IR region practically for any ionization degree. 
Second, there is always a metastable state in this configuration with long
enough life time to be used as a clock state. The latter can be understood
using simple arguments. The fine structure of the $4f$ states in the highly
charged ions is large and the lowest states of the $4f^{12}$
configuration can be 
considered as the states of the two-hole states of the $4f^2_{7/2}$
relativistic configuration.  The 
states of this configuration can have the total angular momentum $J=6,4,2,0$.
According to the Hund's rules, the 
 $J=6$ state is the ground state and $J=4$ state is the first
excited state.  
The excited state can only decay to the ground state via electric quadrupole
transition. This makes it a very long living state, suitable for atomic clock.

Similar consideration holds for any ions with the $nl^2$ ($l=1,2,3$)
two-electron or two-hole ground state configuration, e.\ g.\ $4f^2$,
$nd^8$ and $nd^2$ ($n=3,4,5$), and $np^4$ ($n=2,3,4,5,6$). However,
the radiative width of the states tends to increase with the
decreasing value of 
the total angular momentum. For example, the width of the
states of the $4f^2$ configuration ($4f^2_{5/2}$ electron states) is
roughly one order of magnitude larger than the width of the states of
the $4f^{12}$ configuration ($4f^2_{7/2}$ hole states). The width of
the states of the $4d^8$ configuration ($4d^2_{5/2}$ hole states) is
close to those of the $4f^2$ configuration. But the width of the
states of the $4d^2$ configuration ($4d^2_{3/2}$ electron states) is
larger again. For this reason in present paper we only consider the
states of the $4f^{12}$ configuration which can be used to build the
most accurate HCI optical clocks.

In this paper, we study ions which have electron configuration of
palladium or cadmium plus twelve $4f$ electrons: [Pd]$5s^24f^{12}$ or
[Pd]$4f^{12}$. The [Pd]$4f^{12}$
configuration is the ground state configuration for all ions starting from
Re$^{17+}$ which have nuclear charge $Z \geq 75$ and  degree of
ionization $Z_i = Z-58$. These are not the only ions which have the
$4f^{12}$ configuration in the ground state. For example, neutral
erbium has the [Xe]$4f^{12}6s^2$ ground-state configuration, Er~III
ions has the [Xe]$4f^{12}$ ground-state configuration, etc. Many
properties of these ions and neutral erbium are very similar to those
of the HCIs. However, HCIs are naturally more suitable for accurate
time-keeping because of their smaller electronic-cloud size and
thereby suppressed  couplings and lower sensitivity to external
perturbations. 

\begin{table}
\caption{Properties of clock transitions in even-even isotopes
  (nuclear spin $I=0$) of 
  highly charged ions with the $4f^{12}$ ground-state configuration
  of valence electrons. The complete ground state configuration is
  [Pd]$5s^24f^{12}$ for Hf$^{12+}$ and  W$^{14+}$ and [Pd]$4f^{12}$
  for other ions. $\Delta E$ is the energy interval between the ground
  and the excited 
  clock states, $\lambda$ is corresponding wavelength, $\Gamma$ is the
  radiative 
  width of the clock state, $\tau$ is its lifetime, and $Q$ is the quality
  factor ($Q = \omega/\Gamma$). Numbers in square brackets represent
  powers of 10.}  
\label{t:isotope}
\begin{ruledtabular}
\begin{tabular}{crrrrcc}
$Z$ & \multicolumn{1}{c}{Ion} &\multicolumn{1}{c}{$\Delta E$}  
    & \multicolumn{1}{c}{$\lambda$} &\multicolumn{1}{c}{$\Gamma$}  
    & \multicolumn{1}{c}{$\tau$} &\multicolumn{1}{c}{$1/Q$} \\  
    &     & \multicolumn{1}{c}{cm$^{-1}$} & \multicolumn{1}{c}{nm} & 
 \multicolumn{1}{c}{$\mu$Hz} & \multicolumn{1}{c}{days} & \\ 
\hline
72 & $^{180}$Hf$^{12+}$ &   8555 &  1168 &   9.5 &   4.6 &  3.7[-20] \\
74 & $^{184}$W$^{14+}$  &   9199 &  1087 &   9.6 &   4.6 &  3.5[-20] \\
76 & $^{192}$Os$^{18+}$ &   9918 &  1008 &  13.6 &   3.2 &  4.6[-20] \\
78 & $^{194}$Pt$^{20+}$ &  10411 &   960 &  13.5 &   3.3 &  4.3[-20] \\
80 & $^{202}$Hg$^{22+}$ &  10844 &   922 &  13.2 &   3.4 &  4.1[-20] \\
82 & $^{208}$Pb$^{24+}$ &  11257 &   888 &  12.8 &   3.4 &  3.8[-20] \\
84 & $^{208}$Po$^{26+}$ &  11624 &   860 &  12.3 &   3.6 &  3.5[-20] \\
88 & $^{226}$Ra$^{30+}$ &  12275 &   814 &  11.2 &   3.9 &  3.1[-20] \\
90 & $^{232}$Th$^{32+}$ &  12567 &   795 &  10.7 &   4.1 &  2.8[-20] \\
92 & $^{238}$U$^{34+}$  &  12841 &   778 &  10.2 &   4.4 &  2.6[-20] \\
\end{tabular}
\end{ruledtabular}
\end{table}

Table \ref{t:isotope} lists  relevant properties of most abundant stable
even-even isotopes of the HCIs which have the $4f^{12}$ configuration
of the ground state. All enumerated isotopes have vanishing nuclear spin.
Numerical calculations were carried out
with the version of the configuration interaction method described in
\cite{DF08a,DF08b}.

The probability of the electric quadrupole (E2) transition is given by (we use
atomic units: $\hbar=1$, $m_e=1$, $|e|=1$)
\begin{equation}
\Gamma_e = \frac{1}{15}\alpha^5\omega_{ab}^5\frac{\langle e||E2||
  g\rangle^2}{2J_e + 1}. \label{eq:pe2}
\end{equation}
Here $g$ is the ground state and $e$ is the excited metastable state, $\alpha
= 1/137.36$ is the fine structure constant, $\omega_{ge}$ is the frequency of
the clock transition. Typical value of frequency for the considered
transitions is 
$\omega_{eg} \sim 0.1$ a.u., the amplitude of the E2 transition 
$\langle e||E2||g\rangle \ll 1$ a.u. , $J_e=4$. This leads to
$\Gamma_e \sim 10^{-21}$ a.u. ($\sim 10 \mu$Hz) and $Q=\omega_{eg}/\Gamma_e
\sim 10^{20}$ (see Table~\ref{t:isotope}).

The main factors which may affect the performance of the clocks are
quadrupolar shift, blackbody radiation (BBR), static and dynamic Stark
shifts, Zeeman shift and the effect of micromotion. All these and
additional effects 
were considered in detail in our previous paper~\cite{IonClock}. Here
we focus on the electric quadrupole shift using an alternative approach.

\paragraph{Electric quadrupole shift.}
%\section{Electric quadrupole shift}

\begin{table}
\caption{Electric quadrupole moments of the ground ($J=6$) and excited
  ($J=4$) clock states for ions from Hf$^{12+}$ to U$^{34+}$ and the
  amplitude of the E2 transition between the states. The numbers are
  in atomic units.}
\label{t:Q}
\begin{ruledtabular}
\begin{tabular}{ccccc}
Ion & $Q_6$ & $Q_4$ & $Q_6/Q_4$ & $\langle 6||E2||4\rangle$ \\
\hline
Hf$^{12+}$ &   0.2276 &  -0.0132 &    -17.2 &   0.3240 \\
W$^{14+}$  &   0.1879 &  -0.0137 &    -13.7 &   0.2715 \\
Os$^{18+}$ &   0.1837 &  -0.0151 &    -12.1 &   0.2680 \\
Pt$^{20+}$ &   0.1611 &  -0.0141 &    -11.4 &   0.2364 \\
Hg$^{22+}$ &   0.1430 &  -0.0130 &    -11.0 &   0.2106 \\
Pb$^{24+}$ &   0.1282 &  -0.0120 &    -10.7 &   0.1892 \\
Po$^{26+}$ &   0.1158 &  -0.0110 &    -10.5 &   0.1712 \\
Ra$^{30+}$ &   0.0965 &  -0.0093 &    -10.4 &   0.1427 \\
Th$^{32+}$ &   0.0888 &  -0.0086 &    -10.3 &   0.1312 \\
U$^{34+}$  &   0.0821 &  -0.0080 &    -10.3 &   0.1212 \\
\end{tabular}
\end{ruledtabular}
\end{table}

\begin{table*}
\caption{Transitions convenient for use to suppress the electric
  quadrupole shift. $J_1,M_1$ are the total angular momentum and its
  projection for the ground state, $J_2,M_2$ are the total angular
  momentum and its projection for the clock state, $A$ is given by
  (\ref{eq:A}), $c_1=1/(1-A)$, $c_2=A/(A-1)$ ($\omega_0 = c_1\omega_1+c_2\omega_2$).}
\label{t:A}
\begin{ruledtabular}
\begin{tabular}{cccccccc}
$J_1,M_1 - J_2,M_2$ & $J_1,M^{\prime}_1 - J_2,M^{\prime}_2$ 
%$JM - JM$ & $JM - JM$ 
& $A$ & $c_1$ & $c_2$ & $A$ &$c_1$ & $c_2$ \\ 
\hline
&& \multicolumn{3}{c}{$Q_6/Q_4=-17$} &  
 \multicolumn{3}{c}{$Q_6/Q_4=-10$} \\  
  6,2 - 4,0 & 6,5 - 4,4 & -0.9351 & 0.5168 & 0.4832 & -0.9578 & 0.5108 & 0.4892 \\
  6,2 - 4,1 & 6,5 - 4,4 & -0.9494 & 0.5130 & 0.4870 & -0.9846 & 0.5039 & 0.4961\\
  6,2 - 4,2 & 6,5 - 4,4 & -0.9922 & 0.5020 & 0.4980 & -1.0649 & 0.4843 & 0.5157\\
  6,2 - 4,3 & 6,5 - 4,3 & -0.9669 & 0.5084 & 0.4916 & -1.0096 & 0.4976 & 0.5024\\
\end{tabular}
\end{ruledtabular}
\end{table*}

One of the most important systematic effects is the
clock frequency shift due to interaction of ionic quadrupole moments with
the gradients of trapping electric field. In our previous
paper~\cite{IonClock} we suggested using hyperfine structure of the
clock states to suppress the shift. Here we
explore a different approach based on combining transition frequencies
between states of different projections of the total angular
momentum $J$.
 
The coupling of Q-moment to the E-field gradient ${\partial E_z}/{\partial z}$ reads ($z$ is the  quantization axis determined by externally applied B-field)
\begin{equation}
H_Q = -\frac{1}{2}Q\frac{\partial E_z}{\partial z}.
\label{eq:hq}
\end{equation}  
%where $Q$ is the quadrupole moment of the specific ionic state. 
The quadrupole moment $Q$ of the atomic state is defined conventionally as
twice the expectation value in the stretched state
\begin{equation}
Q_J=2\langle nJM=J|Q_0|nJM=J\rangle.
\label{eq:q}
\end{equation}
Calculated values of $Q$ for the ground $Q_6$ and excited $Q_4$ states
are compiled in Table~~\ref{t:Q}. 

%Taking $\frac{\partial E_z}{\partial z} \approx 10^3 \ {\rm V}/{\rm cm}^2$ as
%for Hg$^+$ \cite{Hg+} and 

Typical values of the gradient $\partial \mathcal{E}_{z} / \partial z \approx 10^8 \, \mathrm{V}/\mathrm{m}^2$ and Q-moments from Table~\ref{t:Q} one can get
e.g.,  for Os$^{18+}$
\begin{equation}
\left(\frac{\Delta\nu}{\nu}\right) \sim 10^{-16},
\label{eq:qos}
\end{equation}
which is well above the sought fractional accuracy level.
%One way to supress this shift is by using the hyperfine structure of
%ions~\cite{IonClock}. Here we will use instead the transitions between
%states of different projection of the total angular momentum $J$.

The Q-induced energy shift for a state with total angular momentum $J$ and
its projection $J_z=M$ reads
\begin{equation}
  \delta E_{JM} ~\sim \frac{3M^2-J(J+1)}{3J^2-J(J+1)}Q_J\frac{\partial
    E_z}{\partial z} \equiv
  C_{JM}Q_J\frac{\partial E_z}{\partial z}.
\label{eq:at}
\end{equation}
%where $Q$ is the electric quadrupole moment of the state and $\nabla
%E$ is the gradient of electric field.
Clock frequency of the transition between two states $J_1,M_1$ and $J_2,M_2$ can
be expressed as
\begin{equation}
  \omega = \omega_0 + \left(C_{J_1,M_1}Q_{J_1} +
  C_{J_2,M_2}Q_{J_2}\right) \frac{\partial E_z}{\partial z},
\label{eq:qs}
\end{equation}
where $\omega_0$ is the unperturbed clock frequency. The uncertainty due to the electric quadrupole
shift can be eliminated if two transitions between states with
different projections $M$ are considered. Indeed, using the expression
(\ref{eq:qs}) for two transitions $J_1,M_1 - J_2,M_2$ with frequency
$\omega_1$ and $J_1,M^{\prime}_1 - J_2,M^{\prime}_2$ with frequency
$\omega_2$ one can find the unperturbed frequency $\omega_0$:
\begin{equation}
  \omega_0 = \frac{\omega_1 -A\omega_2}{1-A},
\label{eq:w0}
\end{equation}
where
\begin{equation}
  A = \frac{C_{J_1,M_1}(Q_{J_1}/Q_{J_2}) +
    C_{J_2,M_2}}{C_{J_1,M^{\prime}_1}(Q_{J_1}/Q_{J_2}) +
    C_{J_2,M^{\prime}_2}}.
\label{eq:A}
\end{equation}
Expressions (\ref{eq:w0}) and (\ref{eq:A}) do not depend on the E-field
gradient. 

The uncertainty due to the quadrupole
shift can be eliminated if quadrupole moments of both states are
known. To be precise, we only need to know their ratio. In the approximation
of pure two-hole configuration $4f^2_{7/2}$ this ratio can be found
analytically. The quadrupole moment for a state with total angular
momentum $J$ is given by
\begin{eqnarray}
 && Q_J = - (2J+1) \label{eq:QJ} \\
&& \times \left( \begin{array}{rrr} J &2&J\\-J&0&J \end{array} \right)
\left\{ \begin{array}{ccc} J &2&J\\j&j&j \end{array} \right\} 
\langle j||E2||j \rangle. \nonumber
\end{eqnarray}
Here $J$ = 6 or 4 and $j=7/2$. It follows from (\ref{eq:QJ}) that
$Q_6/Q_4=-11$. For this value of the ratio, the quadrupole shift is
cancelled out by simple averaging of the frequencies of the
two transitions $M_1=2,M_2=3$ and $M^{\prime}_1=5,M^{\prime}_2=3$:
\begin{equation}
  \omega_0 =(\omega_1+\omega_2)/2,
\label{eq:w05}
\end{equation}
where
\begin{eqnarray}
  && \omega_1 = E(J=4,M=3) - E(J=6,M=2)   \nonumber \\
  && \omega_2 = E(J=4,M=3) - E(J=6,M=5)   \label{eq:w1w2}
\end{eqnarray}
The true value of the $Q_6/Q_4$ ratio may differ from the approximate value of $-11$
(see Table \ref{t:Q}) mostly due to the admixture of the
$4f_{7/2}4f_{5/2}$ configuration. If this ratio is known (from
calculations or measurements) the use of (\ref{eq:w0}) and
(\ref{eq:A}) ensures accurate cancellation of the quadrupole
shift. Table~\ref{t:A} lists some convenient E2-allowed transitions. 

Note that the computed ratio $Q_6/Q_4$ varies relatively little from ion to
ion. For all ions with the [Pd]$4f^{12}$ configuration of the ground
state (from Os$^{18+}$ to U$^{34+}$) it is within 10\% of the
analytical value of $-11$. For all these values the use of simplest
case (\ref{eq:w05}), (\ref{eq:w1w2}) leads to at least two orders of
magnitude suppression of the quadrupole shift.

\paragraph{Other systematics}
%\section{Other systematics}

Systematic effects which can affect the performance of the ionic clocks
with the $4f^{12}$ configuration of the ground state were studied in
detail in our previous work~\cite{IonClock}. In addition to electric quadrupole shift
considered above, they include frequency shift due to
black-body radiation (BBR), Zeeman shift, Doppler effect and
gravity. Actual estimations were done for the Os$^{18+}$, Bi$^{25+}$,
and U$^{34+}$ ions and discussed in detail for the Bi$^{25+}$ ion. It
was clear from the analysis that parameters of the ions vary
relatively little from one ion to another and the analysis performed
in~\cite{IonClock} is valid for all ions considered in the present paper. 

% We do not need it here
%The fractional frequency shift due to room-temperature BBR is estimated to be $\sim
%6\times 10^{-21}$. The BBR shift in cryogenic Paul trap operating at
%the liquid helium temperature ($\sim 4K$) would be further suppressed
%by many orders of magnitude.

Compared to Ref. ~\cite{IonClock}, the absence of hyperfine structure in presently considered isotopes 
modifies analysis of second-order Zeeman shifts.
%Clock frequencies are affected by magnetic fields. The first-order Zeeman
%shift can be eliminated by averaging the measurements over two virtual 
%clock transitions with opposite g-factors. 
The second-order AC Zeeman
shift was estimated in \cite{IonClock} assuming the value of the
magnetic field $B_{AC}=5\times 10^{-8}$T measured in the Al$^+$/Be$^+$
trap \cite{AlBe} and found to be $4\times 10^{-20}$. Note however,
that the second-order Zeeman shift is strongly enhanced in ions
considered in \cite{IonClock} due to small energy intervals between
states of the hyperfine structure multiplet. In present paper we focused on
 ions lacking hyperfine structure. This means that the
second-order Zeeman shift is further suppressed for these ions by
several orders of magnitude. This is important advantage for using
these ions.
 
It was shown in \cite{IonClock} that all other systematic effects
produce fractional frequency shift which is below the value of $10^{-19}$. We anticipate that due 
to simplified level structure of nuclear-spin-zero isotopes, the present work may provide a simpler and potentially more accurate route to HCI-based clocks that can carry out metrology at the 19th significant figure.

\section*{Acknowledgments}

The authors are grateful to G. Gribakin for useful discussions.
The work was supported in part by the Australian Research Council and the
U.S. National Science Foundation.

\end{document}